\documentclass{iau} 
\usepackage{graphicx}

\title[Protostellar Population in the CMZ] 
{Deeply Embedded Protostellar Population in the Central Molecular Zone Suggested by H$_2$O Masers and Dense Cores}

\author[Xing Lu et al.]   
{Xing Lu$^{1,2
\thanks{Present address: National Astronomical Observatory of Japan, 2-21-1 Osawa, Mitaka, Tokyo 181-8588, Japan.}}$,
 Qizhou Zhang$^2$,
 Jens Kauffmann$^3$,
 Thushara Pillai$^3$,
 Steven~N.~Longmore$^4$,
 J.~M.~Diederik Kruijssen$^5$,
 \and Cara Battersby$^2$}

\affiliation{$^1$School of Astronomy and Space Science, Nanjing University, Nanjing,
Jiangsu 210023, China \\[\affilskip]
$^2$Harvard-Smithsonian Center for Astrophysics, Cambridge, MA 02138, USA \\[\affilskip]
$^3$Max Planck Institut f\"ur Radioastronomie, Auf dem H\"ugel 69, D-53121 Bonn, Germany \\[\affilskip]
$^4$Astrophysics Research Institute, Liverpool John Moores University, 146 Brownlow Hill, Liverpool L3 5RF, UK \\[\affilskip]
$^5$Zentrum f\"ur Astronomie der Universit\"at Heidelberg, Astronomisches Rechen-Institut, M\"onchhofstra{\ss}e 12-14, D-69120 Heidelberg, Germany
}

\pubyear{2016}
\volume{322}  
\jname{The Multi-Messenger Astrophysics of the Galactic Centre}
\editors{Steven Longmore, Geoff Bicknell, \& Roland Crocker, eds.}
\begin{document}

\maketitle

\begin{abstract}
The Central Molecular Zone (CMZ), usually referring to the inner 500 pc of the Galaxy, contains a dozen of massive ($\sim$10$^5$~$M_\odot$) molecular clouds. Are these clouds going to actively form stars like Sgr~B2? How are they affected by the extreme physical conditions in the CMZ, such as strong turbulence? Here we present a first step towards answering these questions. Using high-sensitivity, high angular resolution radio and (sub)millimeter observations, we studied deeply embedded star formation in six massive clouds in the CMZ, including the 20 and 50 km\,s$^{-1}$ clouds, Sgr B1~off (as known as dust ridge clouds e/f), Sgr~C, Sgr~D, and G0.253$-$0.016. The VLA water maser observations suggest a population of deeply embedded protostellar candidates, many of which are new detections. The SMA 1.3 mm continuum observations reveal peaks in dust emission associated with the masers, suggesting the existence of dense cores. While our findings confirm that clouds such as G0.253$-$0.016 lack internal compact substructures and are quiescent in terms of star formation, two clouds (the 20 km\,s$^{-1}$ cloud and Sgr~C) stand out with clusters of water masers with associated dense cores which may suggest a population of deeply embedded protostars at early evolutionary phases. Follow-up observations with VLA and ALMA are necessary to confirm their protostellar nature.
\end{abstract}

\firstsection 
\section{Introduction}

Massive molecular clouds in the Central Molecular Zone (CMZ) of the Galaxy are known to be inactive in star formation (e.g., \cite[Lis et al.~1994]{lis1994}; \cite[Immer et al.~2012]{immer2012}) and the overall star formation rate of the CMZ is estimated to be an order of magnitude lower than expected from the Kennicutt-Schmidt relations (\cite[Longmore et al.~2013]{longmore2013}). One case of inactive star forming clouds in the CMZ is G0.253$-$0.025, which contains $>$10$^5$~$M_\odot$ of molecular gas (\cite[Longmore et al.~2012]{longmore2012}) but one weak H$_2$O maser associated with a dense core has been found so far (\cite[Lis et al.~1994]{lis1994}; \cite[Kauffmann et al.~2013]{kauffmann2013}; \cite[Johnston et al.~2014]{johnston2014}; \cite[Rathborne et al.~2015]{rathborne2015}).

\cite[Kruijssen et al.~(2014)]{kruijssen2014} have proposed that star formation in the CMZ may proceed episodically, in which case it is currently at a low point of star formation cycles. Strong turbulence in the CMZ clouds inhibits collapse of dense gas, hence suppresses star formation in the clouds (\cite[Kruijssen et al.~2014]{kruijssen2014}; \cite[Krumholz \& Kruijssen 2015]{krumholz2015}; \cite[Krumholz et al.~2016]{krumholz2016}). Meanwhile, once gravitational collapse starts, subsequent star formation seems to proceed within normal time scales (\cite[Kruijssen et al.~2015]{kruijssen2015}).

\begin{figure}[!t]
\begin{center}
 \includegraphics[width=6.0in]{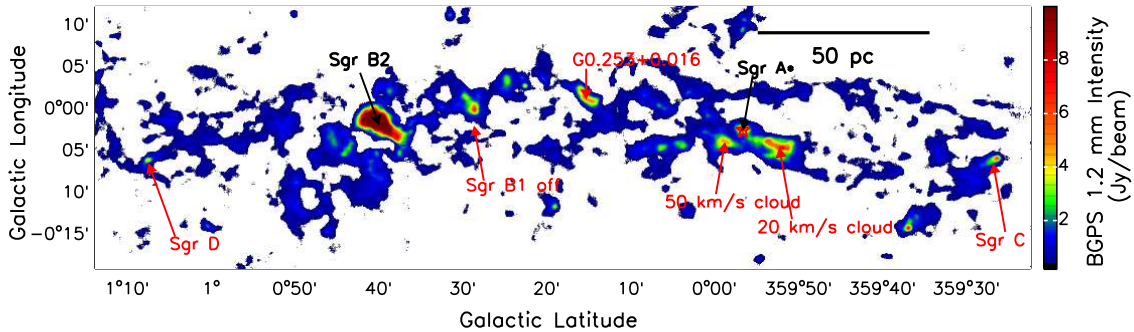} 
 \caption{Overview of the six massive molecular clouds in the CMZ. The background image shows the BGPS 1.2~mm continuum emission (\cite[Ginsburg et al.~2013]{ginsburg2013}). The six clouds are marked by red arrows. Sgr~A* and Sgr~B2, which are not covered in our observations, are also marked by black arrows.}
   \label{fig1}
\end{center}
\end{figure}

To explore the new generation of star formation in the CMZ expected by the episodic star formation model, we started to search for deeply embedded protostellar population at very early evolutionary phases in massive molecular clouds. We selected six clouds with column densities above 10$^{23}$~cm$^{-2}$, hence containing large amount of dense gas and the most likely star forming regions (Fig.\,\ref{fig1}). In \cite[Lu et al.~(2015)]{lu2015b}, we have reported the discovery of a number of protostellar candidates traced by H$_2$O masers and associated dense cores in one of the clouds, the 20 km\,s$^{-1}$ cloud. Here we continued to report our progress on searching for protostellar candidates in all the six clouds.

\section{Results}

We obtained the Karl~G.~Jansky Very Large Array (VLA) 1.3~cm continuum and multiple NH$_3$ inversion line observations, as well as the Submillimeter Array (SMA) 1.3~mm continuum and spectra line observations, of the six clouds. In particular, we used the 1.3~mm continuum, which principally comes from dust emission, to trace dense cores of $\sim$0.1~pc scales, and the 22.2~GHz H$_2$O maser line emission to pinpoint protostellar population.

\begin{figure}[!t]
\begin{center}
 \includegraphics[width=4.4in]{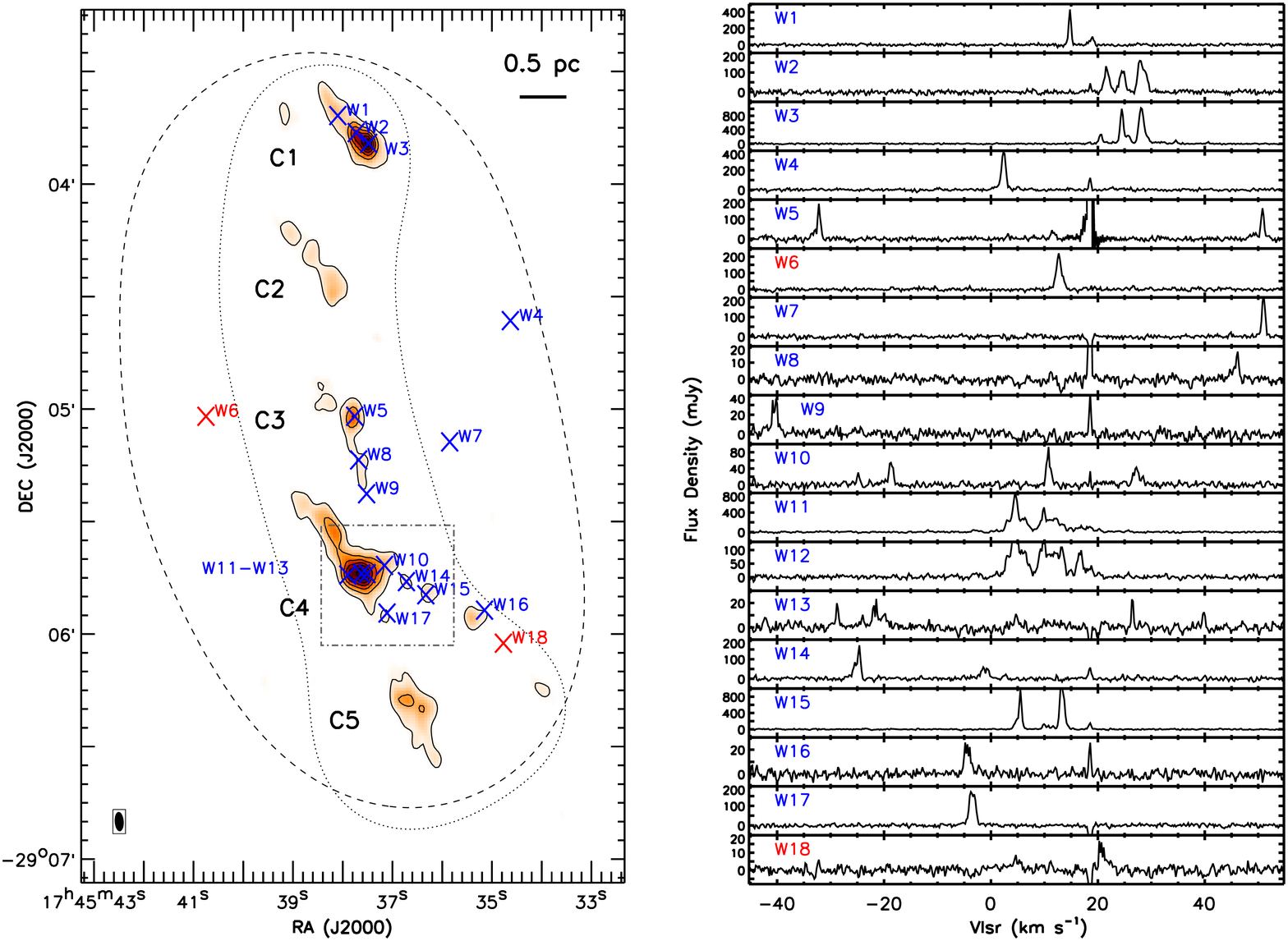} 
 \caption{Detection of H$_2$O masers and dense cores in the 20~km\,s$^{-1}$ cloud (\cite[Lu et al.~2015]{lu2015b}). \textit{Left}: The background image and the contours show the SMA 1.3~mm continuum emission, with levels between 5$\sigma$ and 65$\sigma$ in steps of 10$\sigma$, where 1$\sigma$=3~mJy\,beam$^{-1}$. The crosses show the H$_2$O masers. The two red crosses (W6, W18) are masers with known OH/IR star counterparts. The dotted and green dashed loops show the FHWM of the SMA and VLA primary beams, respectively. \textit{Right}: The spectra of the 18 H$_2$O masers. 
}
   \label{fig2}
\end{center}
\end{figure}

First, we identified H$_2$O masers in the six clouds. In the end, 50 H$_2$O masers are found, among which $\sim$30 are new detections thanks to the high sensitivity of the VLA observations. H$_2$O masers have been detected in both star forming regions (\cite[Elitzur et al.~1989]{elitzur1989}) and envelopes of AGB stars (\cite[Sjouwerman \& van Langevelde 1996]{sjouwerman1996}). We compared them with the AGB star catalogues in e.g., \cite[Lindqvist et al.~(1992)]{lindqvist1992}, \cite[Sevenster et al.~(1997)]{sevenster1997}, and \cite[Sjouwerman et al.~(1998, 2002)]{sjouwerman1998}, and excluded three masers with known AGB star counterparts in the following analyses. In two of the clouds, the 20~km\,s$^{-1}$ cloud and Sgr~C, we found clusters (more than 10 within 1~pc) of strong H$_2$O masers (fluxes$>$1~Jy) that may trace active star formation. The former cloud is shown as an example in Fig.~\ref{fig2}. G0.253$-$0.016, Sgr~B1~off (as known as clouds e/f), Sgr~D, and the 50~km\,s$^{-1}$ cloud also present a few H$_2$O masers, some of which are new detections.

Then, we searched for dense cores with a visual inspection of the 1.3~mm continuum maps. In Sgr~D, a prominent H\,{\sc ii} region dominates the continuum emission; the other five clouds do not show strong free-free emission from H\,{\sc ii} regions and the 1.3~mm continuum can be used to trace dust emission from dense cores. We found a number of dense cores in the 20~km\,s$^{-1}$ cloud and Sgr~C. Their masses, as derived from dust emission, range from $\sim$100~$M_\odot$ to $\gtrsim$10$^3$~$M_\odot$. We used dense gas tracers, e.g., H$_2$CO and CH$_3$CN, to derive the linewidth in the dense cores, and confirmed that most of the dense cores are gravitationally bound through a virial analysis. It is interesting to note that most of the H$_2$O masers we found are coincident with these dense cores in both spatial coordinates and velocities. In the other three clouds, G0.253$-$0.016, Sgr~B1~off, and the 50~km\,s$^{-1}$ cloud, we found a few dense cores which only constitute a small fraction of the cloud masses.

In summary, we found two outstanding cases, the 20~km\,s$^{-1}$ cloud and Sgr~C, which present clusters of H$_2$O masers and dense cores. Meanwhile, previous observations have shown that these two clouds do not have prominent H\,{\sc ii} regions (\cite[Downes et al.~1979]{downes1979}; \cite[Kendrew et al.~2013]{kendrew2013}). Therefore, the two clouds are likely at very early evolutionary phases, while the H$_2$O masers associated with the dense cores likely trace deeply embedded star formation activities. Clouds including G0.253$-$0.016, Sgr~B1~off, and the 50~km\,s$^{-1}$ cloud also present a few H$_2$O masers and dense cores, but overall they remain to be inactive in star formation. Sgr~D presents distinctly different properties with the other clouds. One potential explanation for this is that it may not be located in the CMZ.

\section{Discussions}

As we stressed above, the H$_2$O masers and dense cores only represent protostellar `candidates', since there are still debates on the nature of the H$_2$O masers found in the CMZ. To verify their protostellar nature, it is necessary to obtain supplementary observations of, e.g., class II CH$_3$OH masers, centimeter continuum, or hot core tracers such as CH$_3$CN. These observations would be useful for confirming the existence of deeply embedded star formation in forms of ultra-compact H\,{\sc ii} regions or hot molecular cores. Therefore, future high sensitivity, high angular resolution VLA and ALMA observations of these protostellar candidates are critical.

Once confirmed as protostellar, these masers and dense cores could be used to estimate star formation rates of the clouds in a short time scale of 10$^5$ years. This time scale corresponds to the characteristic evolutionary time before (compact) H\,{\sc ii} regions emerge (\cite[Gerner et al.~2014]{gerner2014}) as well as the typical life time of the class 0 phase of protostars (\cite[Dunham et al.~2014]{dunham2014}). On a qualitative level, the 18 protostellar candidates in the 20~km\,s$^{-1}$ cloud seems to suggest a more active star formation in the last 10$^5$ years than that traced by a single H\,{\sc ii} region in a time scale of 10$^6$ years (\cite[Peters et al.~2010]{peters2010}). This may suggest an increase of star formation activities in the last 10$^5$ years than in the last 10$^6$ years in this cloud.

\end{document}